\documentclass[onecolumn,english,smallextended]{svjour3}
\usepackage{amsmath}
\usepackage{graphicx}
\usepackage{amssymb}

\usepackage{babel}

\begin{document}

\title{Gravitational sources induced by exotic smoothness}

\author{Torsten Asselmeyer-Maluga \and Carl H. Brans}

\institute{T. Asselmeyer-Maluga \at Aerospace Center (DLR), Berlin \\ \email{torsten.asselmeyer-maluga@dlr.de}
\and C.H. Brans \at Loyola University, New Orleans\\ http://www.loyno.edu/~brans \\ \email{brans@loyno.edu}
}

\date{Received: date / Accepted: date}

\maketitle
\begin{abstract}
In this paper we construct a coordinate atlas in an exotic $\mathbb{R}^{4}$
using Bizaca's construction. The main source for such an atlas is
the handle body decomposition of a Casson handle, which of course
is an infinite, but periodic, process. The immersion of the end-periodic
manifold into $\mathbb{R}^{4}$ is directly related to the exoticness
of the $\mathbb{R}^{4}$ and also gives rise naturally to a spinor
field. Thus we obtain the interesting result that the simplest exotic
$\mathbb{R}^{4}$ generates an extra spinor field by exoticness. 
\end{abstract}




\section{Introduction}

The existence of exotic (non-standard) smoothness on topologically
simple 4-manifolds such as exotic $\mathbb{R}^{4}$ or $S^{3}\times\mathbb{R},$
has been known since the early eighties but the use of them in physical
theories has been seriously hampered by the absence of finite coordinate
presentations. However, the work of Bizaca and Gompf \cite{BizGom:96}
provides a handle body representation of an exotic $\mathbb{R}^{4}$
which can serve as an infinite, but periodic, coordinate representation.

Thus we are looking for the decomposition of manifolds into small
non-trivial, easily controlled objects (like handles). As an example
consider the 2-torus $T^{2}=S^{1}\times S^{1}$ usually covered by
at least 4 charts. However, it can be also decomposed using two 1-handles
$D^{1}\times D^{1}$ attached to the $0-$handle $D^{0}\times D^{2}=D^{2}$
along their boundary $\partial D^{2}=S^{1}$ via the boundary component
of the 1-handle $\partial D^{1}\times D^{1}=S^{0}\times D^{1}$, the
disjoint uinon of two lines $S^{0}\times D^{1}=D^{1}\sqcup D^{1}$.
Finally one has to add a 2-handle $D^{2}\times D^{0}$ to get the
closed manifold $T^{2}$. Every 1-handle can be covered by (at least)
two charts and finally we recover the covering by 4 charts. Both pictures
are equivalent but the handle picture has one simple advantage: it
reduces the number of fundamental pieces of a manifold and of the
transition maps. The gluing maps of the handles can be seen as a generalization
of transition maps. Then the handle picture presents only the most
important of these gluing or transition maps, omitting the trivial
transition maps.

In this paper we will present such a coordinate representation, albeit
infinite, of an exotic $\mathbb{R}^{4}$ based on the handle body
decomposition of Bizaca and Gompf. We suggest that one of the consequences
of this approach would be to suggest a positive answer for the Brans
conjecture, that exotic smoothness serves as an additional gravitational
source as a spinor field naturally arising from the handlebody construction.
The compact case was worked out in \cite{AsselmeyerRose2010}.

\section{Construction of exotic $\mathbb{R}^{4}$}

Our model of space-time is the non-compact space topological $\mathbb{R}^{4}$.
The results can be easily generalized for other cases such as $S^{3}\times\mathbb{R}$.

\subsection{Handle decomposition and Casson handle\label{sub:Handle-decomposition-Casson-handle}}

Every 4-manifold can be decomposed using standard pieces such as $D^{k}\times D^{4-k}$,
the so-called $k$-handle attached along $\partial D^{k}\times D^{4-k}$
to the $0-$handle $D^{0}\times D^{4}=D^{4}$. In the following we
need two possible cases: the 1-handle $D^{1}\times D^{3}$ and the
2-handle $D^{2}\times D^{2}$. These handles are attached along their
boundary components $S^{0}\times D^{3}$ or $S^{1}\times D^{2}$ to
the boundary $S^{3}$ of the $0-$handle $D^{4}$ (see \cite{GomSti:1999}).
The attachment of a 2-handle is defined by a map $S^{1}\times D^{2}\to S^{3}$,
the embedding of a circle $S^{1}$ into the 3-sphere $S^{3}$, i.e.
a knot. This knot into $S^{3}$ can be thickened (or a knotted solid
torus). The important fact for our purposes is the freedom to twist
this knotted solid torus via Dehn twist. The (integer) number of these
twists (with respect to the orientation) is called the framing number
or the framing. Thus the gluing of the 2-handle on $D^{4}$ can be
represented by a knot or link together with an integer framing. The
simplest example is the unknot with framing $\pm1$ is the complex
projective space $\mathbb{C}P^{2}$ or with reversed orientation $\overline{\mathbb{C}P}^{2}$,
respectively.

\begin{figure}
\includegraphics{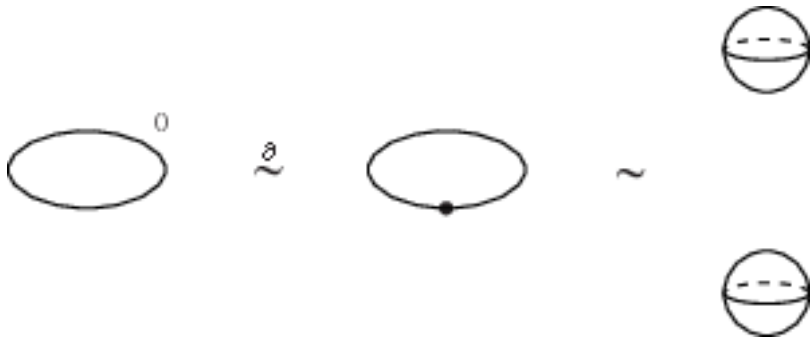}

\caption{Surgering a 2-handle to a 1-handle (The symbol $\partial$ indicates
the diffeomorphism between the boundaries of the corresponding 4-manifolds)
\label{fig:Surgering-1-handle}}

\end{figure}

The 1-handle will be glued by the map of $S^{0}\times D^{3}\to S^{3}$
represented by two disjoint solid 2-spheres $D^{3}$. Akbulut \cite{AkbKir:79}
introduced another description. He observed that a 1-handle is something
like a surgered 2-handle with a fixed framing. The notation in this
figure represents erasing the framing coefficient of the unknot by
putting a dot on it (see figure \ref{fig:Surgering-1-handle}). We
remark that some of the figures are a redrawing of pictures in \cite{GomSti:1999}.

In detail, the procedure can be described as follows. The main observation
is that the union of one 0-handle and $m$ 1-handles ($\approx\natural_{m}S^{1}\times D^{3}$
with $\natural_{m}$ as $m$-times boundary connected sum, see appendix
A) has the same boundary as $m$ 2-handles (0-framed) ($\approx\natural_{m}S^{2}\times D^{2}$)
to an $m$-component unlink (i.e., the boundary of an embedding in
$S^{3}$ of $m$ disjoint disks). In fact, the latter 4-manifold contains
a canonical collection of $m$ (uniquely) framed 2-spheres $S^{2}\times\{*\}\subset S^{2}\times D^{2}$,
and surgery on these framed spheres gives back $\natural_{m}S^{1}\times D^{3}$.
Thus, an $m$-component unlink with a dot on each component is the
same as $m$ 1-handles.

\begin{figure}
\begin{centering}
\includegraphics[scale=0.8]{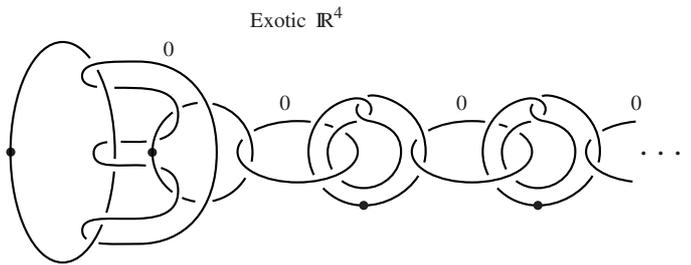}
\par\end{centering}

\caption{Handle decomposition of Bizacas exotic $\mathbb{R}^{4}$\label{fig:Handle-decomposition-exoticR4}}

\end{figure}

Now we are ready to present the handle body decomposition of an exotic
$\mathbb{R}^{4}$ by Bizaca in Fig. \ref{fig:Handle-decomposition-exoticR4}.

\begin{figure}
\begin{centering}
\includegraphics{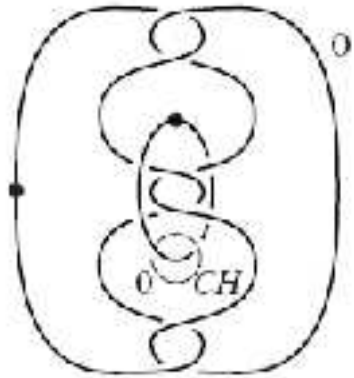}
\par\end{centering}

\caption{Principal picture of an exotic $\mathbb{R}^{4}$\label{fig:exotic-R4}}

\end{figure}

It is very important to notice that the exotic $\mathbb{R}^{4}$ is
the \textbf{interior} of the given handle body (since the handle body
has a non-null boundary). The construction can be divided into two
parts represented in the figure \ref{fig:exotic-R4}. %

\begin{figure}
\begin{centering}
\includegraphics{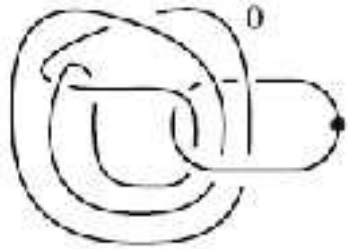}
\par\end{centering}

\caption{Akbulut cork \label{fig:Akbulut-cork}}

\end{figure}

The first part is known as the Akbulut cork represented by figure
\ref{fig:Akbulut-cork}. %

\begin{figure}
\begin{centering}
\includegraphics[scale=0.75]{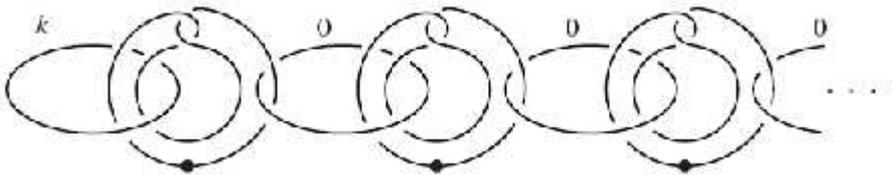}
\par\end{centering}

\caption{Simplest Casson handle \label{fig:Simplest-Casson-handle}}

\end{figure}

In the appendix B we will give a short description of the Akbulut
cork and its meaning for the smoothness of 4-manifolds. The second
part is the Casson handle $CH$ where we use the simplest example
(see figure \ref{fig:Simplest-Casson-handle}). %

\begin{figure}
\includegraphics[scale=0.75]{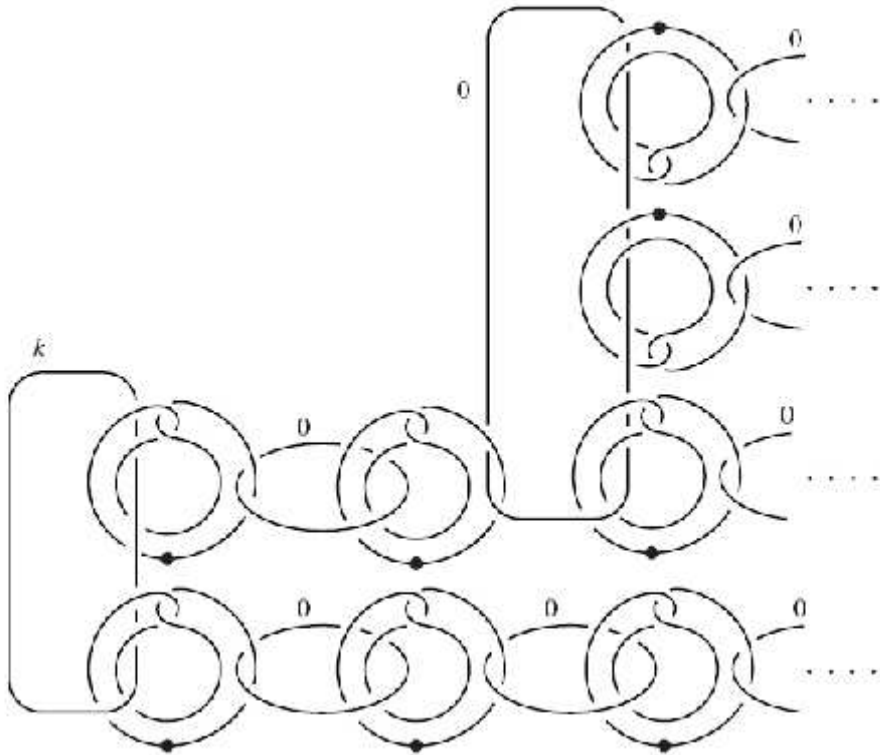}

\caption{More complex Casson handle \label{fig:More-complex-CH}}

\end{figure}

Start with the construction of the Akbulut cork $A$ as a contractible
4-manifold with boundary the homology 3-sphere $\Sigma(2,5,7)$. This
homology 3-sphere is given by the set\[
\Sigma(2,5,7)=\left\{ x,y,z\in\mathbb{C}\,|\, x^{2}+y^{5}+z^{7}=0\,,\,|x|^{2}+|y|^{2}+|z|^{2}=1\right\} \]
 Now it is easy to define the interior $int(A)$ of the cork as the
set\[
int(A)=\left\{ x,y,z\in\mathbb{C}\,|\, x^{2}+y^{5}+z^{7}=0\,,\,|x|^{2}+|y|^{2}+|z|^{2}<1\right\} \]
 This set is a smooth manifold which can be covered by a finite number
of charts. But the smoothness structure of the exotic $\mathbb{R}^{4}$
depends mostly on the Casson handle. If we take (instead of the simplest
handle in Fig.\ref{fig:Simplest-Casson-handle}) the more complex
Casson handle in Fig. \ref{fig:More-complex-CH} then we obtain another
exotic $\mathbb{R}^{4}$ non-diffeomorphic to the previous one. %

Now consider the Casson handle and its construction in more detail.
Briefly, a Casson handle $CH$ is the result of attempts to embed
a disk $D^{2}$ into a 4-manifold. In most cases this attempt fails
and Casson \cite{Cas:73} looked for a substitute, which is now called
a Casson handle. Freedman \cite{Fre:82} showed that every Casson
handle $CH$ is homeomorphic to the open 2-handle $D^{2}\times\mathbb{R}^{2}$
but in nearly all cases it is not diffeomorphic to the standard handle
\cite{Gom:84,Gom:89}. The Casson handle is built by iteration, starting
from an immersed disk in some 4-manifold $M$, i.e. a map $D^{2}\to M$
with injective differential. Every immersion $D^{2}\to M$ is an embedding
except on a countable set of points, the double points. One can kill
one double point by immersing another disk into that point. These
disks form the first stage of the Casson handle. By iteration one
can produce the other stages. Finally consider not the immersed disk
but rather a tubular neighborhood $D^{2}\times D^{2}$ of the immersed
disk including each stage. The union of all neighborhoods of all stages
is the Casson handle $CH$. So, there are two input data involved
with the construction of a $CH$: the number of double points in each
stage and their orientation $\pm$. Thus we can visualize the Casson
handle $CH$ by a tree: the root is the immersion $D^{2}\to M$ with
$k$ double points, the first stage forms the next level of the tree
with $k$ vertices connected with the root by edges etc. The edges
are evaluated using the orientation $\pm$. Every Casson handle can
be represented by such an infinite tree. The Casson handle $CH(R_{+})$
in Fig.\ref{fig:Simplest-Casson-handle} is the simplest Casson handle
represented by the simplest tree $R_{+}$ having one vertex in each
level connected by one edge with evaluation $+$. We will now go into
more detail. The reader not interested in very technical terms can
go directly to the next subsection.

\begin{figure}
\begin{centering}
\includegraphics[scale=0.6]{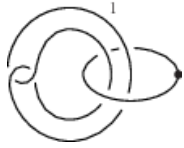}
\par\end{centering}

\caption{Whitehead link \label{fig:Whitehead-link}}

\end{figure}

Each building block of a Casson handle, sometimes called a {}``kinky''
handle, is diffeomorphic to $\natural(S^{1}�D^{3})$ with two attaching
regions. Technically speaking, one region is a tubular neighborhood
of band sums of Whitehead links (see Fig. \ref{fig:Whitehead-link})
connected with the previous block. %

The other region is a disjoint union of the standard open subsets
$S^{1}\times D^{2}$ in $\#S^{1}\times S^{2}=\partial(\natural S^{1}�D^{3})$
(this is connected with the next block). The number of end-connected
sums is exactly the number of self intersections of the immersed two
handle. The simplest Casson handles have $S^{1}\times D^{3}$ as their
building blocks represented by the Fig. \ref{fig:Whitehead-link}.

We attach a Casson handle to the zero\textendash{}handle along the
attaching circle and denote it by $S=D^{4}\cup CH$. Consider a simple
Casson handle, say $CH(R_{+})$, a periodic Casson handle with positive
orientation kinks (see Fig. \ref{fig:Simplest-Casson-handle}). As
shown in \cite{Kato2004}, the Casson handle is a so-called end-periodic
manifold, i.e. a manifold with a periodic structure of building blocks,
discussed in the next subsection. The periodicity of the topological
construction can be naturally translated into the periodicity of the
metric to be imposed on the resulting manifold. The building block
as an open manifold becomes an `open cylindrical' manifold. Then one
connects two attaching regions in a block. The result becomes a cylindrical
manifold on which analysis is already well known. By equipping it
with a suitable weight function, one will apply the generalized Fourier\textendash{}Laplace
transform between complex functions on the cylindrical manifold and
its periodic cover. Thus one is able to construct operators or functions
over $S=D^{4}\cup CH(R(2))$. This method shows that once one obtains
some suitable function spaces on any open manifolds, then the generalized
Fourier\textendash{}Laplace transform (described below) works on
their periodic covers. We will use this observation iteratively. As
described above a Casson handle can be expressed by an infinite tree
with one end point and with a sign $\pm$ on each edge. The next simplest
Casson handle (see Fig. \ref{fig:More-complex-CH}) will be represented
as follows (here we will follow \cite{Kato2004} very closely). Let
$R_{+}$ be the half-line with the vertices $\left\{ 0,1,2,\ldots\right\} $.
We prepare another family of half-lines $\left\{ R_{+}^{i}\right\} _{i=1,2,...}$
assigned with indices. Then we obtain another infinite tree: \[
R(2)=R_{+}\bigcup_{i=1,2,\ldots}R_{+}^{i}\]
 where we connect $i$ in $R_{+}$ with $0$ in $R_{+}^{i}$. For
example one may assign $-$ on $R_{+}$ and $+$ on all $\left\{ R_{+}^{i}\right\} _{i}$.
Then one obtains the corresponding Casson handle $CH(R(2))$. In this
case the building blocks are diffeomorphic to $\hat{\Sigma}_{2}=(S^{1}\times D^{3})\natural(S^{1}\times D^{3})$
along $R_{+}$. $\hat{\Sigma}_{2}$ has three attaching components.
One is $\mu$, the tubular neighborhood of the band sum of two Whitehead
links as before. We will denote the others by $\mu'$ and $\gamma$
, where these represent a generator of $\pi_{1}\left(\hat{\Sigma}_{2}\right)$.
In order to apply Fourier\textendash{}Laplace transform (described
below), one takes end-connected sums twice. Firstly one takes the
end-connected sum between $\mu$ and $\mu'$ as before. The result
is an `open cylindrical' manifold, since there still remains one attaching
region, $\gamma$ . One takes the end-connected sum of this with $CH(R_{+})$
along $\gamma$ . In this manner, one obtains another open manifold,
$\left(\hat{\Sigma}_{2}/\left(\mu\sim\mu'\right)\right)\natural_{\gamma}CH(R_{+})$.
Thus one is again able to construct operators or functions over $S=D^{4}\cup CH(R(2))$.

\subsection{The periodic coordinate patch}

By using the interpretation of the previous subsection \cite{Kato2004},
the Casson handle can be interpreted as an end-periodic manifold.
An end-periodic manifold starts with a compact submanifold, $K,$
with boundary or end $N$ and building blocks $W$ with two ends.
Now we glue the building blocks along a chain to obtain an end-periodic
manifold. In the case of the Casson handle, the compact $K$ is the
Akbulut cork described above and the building block is the tubular
neighborhood of the self-intersecting disk. In the following subsection
we will describe the general approach to end-periodic manifolds and
their analytical properties using extensively Taubes paper \cite{Tau:87}.
Then we will discuss the special end-periodic manifold, the Casson
handle. Finally the coordinate patch is given by the handle decomposition
of the Casson handle.

\subsubsection{Analytical properties of end-periodic manifolds}

\begin{figure}
\begin{centering}
\includegraphics[scale=0.25]{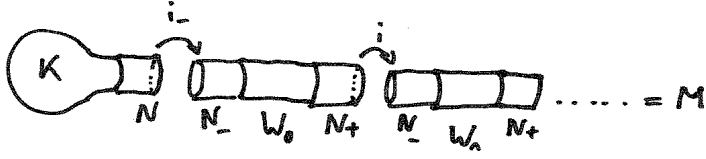}
\par\end{centering}

\caption{\label{fig:end-periodic-manifold}end-periodic manifold (fig. redraw
from \cite{Tau:87})}

\end{figure}

The following definition is very formal and we refer to the Fig. \ref{fig:end-periodic-manifold}.%

A smooth, oriented manifold $M$ is end-periodic if the following
data exists (here we will follow Taubes paper \cite{Tau:87} very
closely):

\begin{enumerate}
\item A smooth, connected, oriented and open manifold $W$ with two ends,
$N_{+}$ and $N_{-}$. $W$ is called the fundamental segment or building
block. Thus, there exists a compact set $C\subset W$ such that $W\setminus C$
is the disjoint union of two nonempty, connected, open sets, $N_{+}$
and $N_{-}$. 
\item Suppose that there is a compact set $C_{+}\subset N_{+}$ such that
$N_{+}\setminus C_{+}$ has two connected components, $N_{++}$ and
$N_{+-}$. Assume that $C_{+}$ is such that $W\setminus C_{+}$ is
the disjoint union of $N_{-}\cup C\cup N_{+-}$ and $N_{++}$. Similarly,
assume that a compact set $C_{-}\subset N_{-}$exists such that $N_{-}\setminus C$
is the disjoint union of two connected components $N_{--}$and $N_{-+}$,
and that $W\setminus C_{-}$ is the disjoint union of $N_{--}$ and
$N_{-+}\cup C\cup N_{+}$. Assume that there is a diffeomorphism $i:N_{+}\to N_{-}$
which is orientation preserving and which takes $N_{++}$to $N_{-+}$
and $N_{+-}$ to $N_{--}$. 
\item An open set $K\subset M$, with one end, $N$. Suppose that a compact
set $C_{0}\subset N$ exists such that $N\setminus C_{0}$ is the
disjoint union of two open sets $N_{0-}$and $N_{0+}$. Assume that
$K\setminus C$ has two components, $\left(K\setminus N\right)\cup N_{0-}$
and $N_{0+}$. Require that there exists a diffeomorphism $i_{-}:N\to N_{-}$
which takes $N_{0-}$to $N_{--}$ and $N_{0+}$ to $N_{-+}$. Require
that $i_{-}$ preserve orientation. 
\item An orientation preserving diffeomorphism $\phi:M\to K\cup_{N}W\cup_{N}W\cup_{N}\cdots$
. Here $K\cup_{N}W$ is obtained from the disjoint union of $K$ and
$W$ by identifying $N\subset K$ with $N_{-}\subset W$ via $i_{-}$.
Also, $W\cup_{N}W$ is obtained from the disjoint union of two copies
of $W$, $W_{1}\cup W_{2}$, by identifying $N_{+}\subset W$, with
$N_{-}\subset W$, via $i_{-}$. (see Fig. \ref{fig:end-periodic-manifold}) 
\end{enumerate}
In particular, there is an identifying map $T:W_{i}\to W_{i+1}$ of
the copies of $W$. With the help of the map $T$ it is possible to
investigate analytical properties of functions. We will follow very
closely the article of Taubes \cite{Tau:87} for the analytical description
of end-periodic manifolds.

First note that the map $i:N_{+}\to N_{-}$ can be used to identify
the two ends of the building block $W$. One obtains $Y=W/i$ having
the $\mathbb{Z}$-fold cover \[
\tilde{Y}=\cdots\cup_{N}W_{-1}\cup_{N}W_{0}\cup_{N}W_{1}\cup_{N}\cdots\]
 as an end-periodic manifold with projection $\pi:\tilde{Y}\to Y$.
Then the end \begin{equation}
End(M)=W_{0}\cup_{N}W_{1}\cup_{N}W_{2}\cup_{N}\cdots\label{eq:end-of-M}\end{equation}
 is a subset of $Y$ and can be identified with $M\setminus(K\setminus N)$.
A vector bundle $E\to M$ is end-periodic if the the map $T$ lifts
to a bundle map $\tilde{T}:E|_{W_{i}}\to E|_{W_{i+1}}$ or if $E|_{End(M)}=\pi^{*}E_{Y}$
where $E_{Y}\to Y$ is a vector bundle and $\pi:End(M)\to Y$ the
projection. Then we obtain the main idea for the analytical description:
\emph{Transform a function over $M$ to a function over $\tilde{Y}$
considered as periodic function}. Thus, it is enough to consider a
function as a smooth, compactly supported section $\psi\in\Gamma_{0}(\tilde{E})$
of a periodic vector bundle $\tilde{E}\to\tilde{Y}$. We define the
\emph{generalized Fourier-Laplace transform (or Fourier-Laplace transform
for short)} of $\psi$ by\begin{equation}
\hat{\psi}_{z}(.)=\sum_{n=-\infty}^{\infty}z^{n}(\tilde{T}^{n}\psi)(.)\label{eq:fourier-laplace}\end{equation}
 with $z=\exp(i\delta)\in S^{1}=\mathbb{C}/\mathbb{Z}=\mathbb{C}^{*}$.
$\hat{\psi}$ defines a smooth section of the vector bundle \[
E_{Y}(z)=\left[\tilde{E}\otimes\mathbb{C}/\mathbb{Z}\right]\]
 over $Y=\tilde{Y}/\mathbb{Z}$, where $\mathbb{Z}$ acts on $\tilde{E}\otimes\mathbb{C}$
via the action sending $1\in\mathbb{Z}$ and $(p,\lambda)\in E\otimes\mathbb{C}$
to $(\tilde{T}p,\, z\lambda)$. The collection \[
E_{Y}=\left\{ E_{Y}(z):z\in\mathbb{C}/\mathbb{Z}\right\} \]
 can be seen as a smooth vector bundle over $Y\times\mathbb{C}/\mathbb{Z}$.
The Fourier-Laplace transform can be inverted as follows: Let $\hat{\eta}$
be any section of $E_{Y}$ over $Y\times\mathbb{C}^{*}$ holomorphic
in $\mathbb{C}^{*}=\mathbb{C}/\mathbb{Z}=S^{1}$. Then, if $s\in(0,\infty)$,
the formula\begin{equation}
(\tilde{T}^{n}\eta)(x)=\frac{1}{2\pi i}\intop_{|z|=s}z^{-n}\hat{\eta}_{z}(\pi(x))\frac{dz}{z}\label{eq:fourier-laplace-inverse}\end{equation}
 for $x\in W_{0}$ and $\pi(x)\in Y$ defines a section of $\tilde{E}$
over $\tilde{Y}$.

\subsubsection{The metric and the periodic coordinate patch}

{We are interested in a smooth metric $g$ over $M$ which can be
seen as a section in the tensor bundle $TM\otimes TM$. One way to
introduce a metric was described in \cite{Kato2004} by using an embedding
of $M$ in some Euclidean space. Here we will use another method to
construct $g$ by a periodic metric $\hat{g}$ on $Y$ giving a metric
on the building block $W$. To reflect the number of the building
block, we have to extend $\hat{g}$ to $Y\times\mathbb{C}^{*}$by
using a metric $\hat{g}_{z}$holomorphic in $z\in\mathbb{C}^{*}=S^{1}$.
From the formal point of view we have\begin{equation}
\hat{g}_{z}(.)=\sum_{n=0}^{\infty}a_{n}z^{n}\cdot\hat{g}(.)\label{eq:periodic-metric-on-YxS1}\end{equation}
 where the coefficient $a_{n}$ represents the building block $W_{n}$
in $End(M)$ (see (\ref{eq:end-of-M})). Without loss of generality
we can choose the coordinates $x$ in $M$ so that the $0$th component
$x_{0}$ is related to the integer $n=[x_{0}]$ via its integer part
$[\:]$. Using the inverse transformation (\ref{eq:fourier-laplace-inverse})
we can construct a smoth metric $g$ in $End(M)$ at the $n$th building
block via\[
(\tilde{T}^{n}g)(x)=\frac{1}{2\pi i}\intop_{|z|=s}z^{-n}\hat{g}_{z}(\pi(x))\frac{dz}{z}\]
 for $x\in End(M)\subset\tilde{Y}$, $s\in(0,\infty)$, $n=[x_{0}]$
and $\pi:\tilde{Y}\to Y$. }

{Let $g_{A}$ be the metric in the interior of the cork $A$. As
discussed above the Casson handle can be interpreted as end-periodic
manifold if the Casson handle is generated by a balanced tree. The
two infinite trees $R_{\pm},R(2)$ in subsection \ref{sub:Handle-decomposition-Casson-handle}
are examples of balanced trees. Using this information together with
the handle body structure of the exotic $\mathbb{R}^{4}$ then we
obtain for the metric $g$ on $M=\mathbb{R}_{\Theta}^{4}$:}

\[
g(x)=\begin{cases}
g_{A}(x) & x\in int(A)\\
(\tilde{T}^{[x_{0}]}g)(x) & x\in End(M)\end{cases}\]
 {which is periodic at the end $End(M)$ of $M$. The end-periodic
structure of the Casson handle induces the periodic coordinate patch
of the exotic $\mathbb{R}^{4}$. }

{But can we make sense of the idea of localizing the exotic smoothness
of the $\mathbb{R}_{\Theta}^{4}$? The work in \cite{Biz:95,BizGom:96}
implies that the Casson handle relative to the attaching circle encoded
in the periodic structure is the main incredient. Thus we have to
analyze the structure of the Casson handle more carefully. The main
ingredient of a Casson handle $CH$ is the immersed disk $D^{2}\hookrightarrow CH$,
i.e. the image is a disk with self-intersections. In the appendix
C we represent such a disk in appropriate coordinates. In the next
section we will investigate this immersed disk representing the attaching
circle of the Casson handle.}

\section{The Brans conjecture}

For the following we assume a trivial Casson handle $CH_{0}=D^{2}\times\mathbb{R}^{2}$
in the standard $M=\mathbb{R}^{4}$ (i.e. having the standard differential
structure) and a non-trivial Casson handle $CH$ in the exotic $N=\mathbb{R}_{\Theta}^{4}$
(i.e. admitting an exotic differential structure). {We assume a metric
$g_{M}$ on $M$ (constructed above) satisfying the source-free Einstein
equation}\begin{equation}
R_{\mu\nu}=0\quad.\label{eq:vacuum-equation}\end{equation}
 The corresponding action is \[
S=\intop_{M}R_{M}\sqrt{g_{M}}d^{4}x\]
 with the scalar curvature $R_{M}$ of $M$. As stated above, the
differential structure depends on the Casson handle $CH$ relative
to the attaching region $\partial CH$. As Bizaca \cite{Biz:94,Biz:94a,Biz:95}
showed the Casson handle will be attached to the Akbulut cork $A$
defined above along a circle. Then the complements $M\setminus CH_{0}$
and $N\setminus CH$ are diffeomorphic. Thus, from the physical point
of view we have the relative action\[
\intop_{M\setminus CH_{0}}R_{M}\sqrt{g_{M}}d^{4}x\]
 but the manifold $M\setminus CH_{0}$ has a boundary $\partial CH_{0}$
and as we learned above the concrete embedding of this boundary (i.e.
the attaching of the Casson handle) determines the differential structure%
\footnote{As Freedman \cite{Fre:82} showed, the interior of every Casson handle
is diffeomorphic to the standard $\mathbb{R}^{4}$, i.e. if one forgets
the attaching of the Casson handle.%
}. But then we need the action with a boundary term\begin{equation}
S(M,CH_{0})=\intop_{M\setminus CH_{0}}R_{M}\sqrt{g_{M}}d^{4}x+\intop_{\partial(M\setminus CH_{0})}K_{CH_{0}}\sqrt{g_{\partial}}d^{3}x\label{eq:action-with-bound}\end{equation}
 where $K_{CH_{0}}$ is the trace of the second fundamental form of
the boundary $\partial CH_{0}$ with metric $g_{\partial}$. {Now
we are looking for the motivation of the action at the boundary. As
shown by York \cite{York1972}, the fixing of the conformal class
of the spatial metric in the ADM formalism leads to a boundary term
which can be also found in the work of Hawking and Gibbons \cite{GibHaw1977}.
Also Ashtekar et.al. \cite{Ashtekar08,Ashtekar08a} discussed the
boundary term in the Palatini formalism. All these discussion suggest
the choice of the following term for a boundary\[
\intop_{\partial(M\setminus CH_{0})}K_{CH_{0}}\sqrt{g_{\partial}}d^{3}x=\intop_{\partial(M\setminus CH_{0})}tr(\theta\wedge R)\]
 by using a frame $\theta$ and the curvature 2-form $R$.} The attaching
region $\partial CH$ can be described as the immersion of $D^{2}\times(0,1)$
into $M=\mathbb{R}^{4}$. In appendix D we describe the spinor representation
of an immersed surface $D^{2}$ in $\mathbb{R}^{3}$ which can be
easily extended to an immersion of the attaching region $D^{2}\times(0,1)$
into $M=\mathbb{R}^{4}$. By using relation (\ref{eq:trace-second-fund-form})
in appendix D we obtain the contribution\begin{equation}
\intop_{\partial(M\setminus CH_{0})}K_{CH_{0}}\sqrt{g_{\partial}}d^{3}x=\intop_{\partial(M\setminus CH_{0})}\psi\gamma^{\mu}D_{\mu}\overline{\psi}\sqrt{g_{\partial}}d^{3}x\label{eq:action-3D}\end{equation}
 to the action (\ref{eq:action-with-bound}). But the spinor representation
has one property: this action functional vanishes if the boundary
is embedded, i.e. has no self-intersections. Thus we obtain only a
contribution to the action \[
\intop_{\partial(N\setminus CH)}\psi\gamma^{\mu}D_{\mu}\overline{\psi}\,\sqrt{g_{\partial}}\, d^{3}x\]
 for the exotic smoothness encoded into the boundary of $N\setminus CH$.

{As described in Appendix E, one can extend the action along the
boundary $\partial(N\setminus CH)$ to the whole 4-dimensional manifold
$N\setminus CH$ to get the action\begin{equation}
S(N,CH)=\intop_{N\setminus CH}(R_{M}+\psi\gamma^{\mu}D_{\mu}\overline{\psi})\sqrt{g_{M}}d^{4}x\label{eq:EH-fermionic-action}\end{equation}
 relative to the attaching region. Thus we obtain the Einstein-Hilbert
action with a source term. }

Now summarize this result. By using Bizacas construction we obtain
a coordinate patch of an exotic $\mathbb{R}^{4}$. The exoticness
is directly related to the attaching region of a Casson handle. That
region can be interpreted as an immersed surface for which we obatin
a representation using a spinor. In \cite{AsselmeyerRose2010} we
also discussed the influence of the other immersed disks in the Casson
handle. The results of this paper can be extended to our case of a
small exotic $\mathbb{R}^{4}$ as well.

Finally we obtain the result: \\
 \emph{Thus in general we obtain the combined action of a spinor field
coupled to the gravitational field. The spinor field is represented
by the complement of an immersed disk in the Casson handle.}

\section*{Appendix}

\subsection*{Appendix A - Connected and boundary connected sum}

Let $M,N$ be two $n$-manifolds with boundaries $\partial M,\partial N$.

The connected sum $M\#N$ is the procedure of cutting out a disk $D^{n}$
from the interior $int(M)\setminus D^{n}$ and $int(N)\setminus D^{n}$
with the boundaries $S^{n-1}\sqcup\partial M$ and $S^{n-1}\sqcup\partial N$,
respectively, and glueing them together along the common boundary
component $S^{n-1}$. The boundary $\partial(M\#N)=\partial M\sqcup\partial N$
is the disjoint sum of the boundaries $\partial M,\partial N$.

The boundary connected sum $M\natural N$ is the procedure of cutting
out a disk $D^{n-1}$ from the boundary $\partial M\setminus D^{n-1}$
and $\partial N\setminus D^{n-1}$ and gluing them together along
$S^{n-2}$ of the boundary. Then the boundary of this sum $M\natural N$
is the connected sum $\partial(M\natural N)=\partial M\#\partial N$
of the boundaries $\partial M,\partial N$.

\subsection*{Appendix B - Akbulut cork and smoothness of 4-manifolds}

Consider the following situation: one has two topologically equivalent
(i.e. homeomorphic), simple-connected, smooth 4-manifolds $M,M'$,
which are not diffeomorphic. There are two ways to compare them. First
one calculates differential-topological invariants like Donaldson
polynomials \cite{DonKro:90} or Seiberg-Witten nvariants \cite{Akb:96}.
But there is another possibility: It is known that one can change
a manifold $M$ to $M'$ by using a series of operations called surgeries.
This procedure can be visualized by a 5-manifold $W$, the cobordism.
The cobordism $W$ is a 5-manifold having the boundary $\partial W=M\sqcup M'$.
If the embedding of both manifolds $M,M'$ in to $W$ induces homotopy-equivalences
then $W$ is called an h-cobordism. Furthermore we assume that both
manifolds $M,M'$ are compact, closed (no boundary) and simply-connected.
As Freedman \cite{Fre:82} showed a h cobordism implies a homeomorphism,
i.e. hcobordant and homeomorphic are equivalent relations in that
case. Furthermore, for that case the mathematicians \cite{CuFrHsSt:97}
are able to prove a structure theorem for such h-cobordisms:\\
 \emph{Let $W$ be a h-cobordism between $M,M'$. Then there are contractable
submanifolds $A\subset M,A'\subset M'$ together with a sub-cobordism
$V\subset W$ with $\partial V=A\sqcup A'$, so that the h-cobordism
$W\setminus V$ induces a diffeomorphism between $M\setminus A$ and
$M'\setminus A'$.} \\
 Thus, the smoothness of $M$ is completely determined (see also \cite{Akbulut08,Akbulut09})
by the contractible submanifold $A$ and its embedding $A\hookrightarrow M$.
One calls $A$, the \emph{Akbulut cork}. According to Freedman \cite{Fre:82},
the boundary of every contractible 4-manifold is a homology 3-sphere.
This theorem was used to construct an exotic $\mathbb{R}^{4}$. Then
one considers a tubular neighborhood of the sub-cobordism $V$ between
$A$ und $A'$. The interior $int(V)$ (as open manifold) of $V$
is homeomorphic to $\mathbb{R}^{4}$. If (and only if) $M$ and $M'$
are homeomorphic, but non-diffeomorphic 4-manifolds then the interior
$int(V)$ is an exotic $\mathbb{R}^{4}$.

\subsection*{Appendix C - Representation of the self-intersecting disk}

Every immersed disk with one double point can be uniquely described
via its boundary. The boundary is a curve with one double point parametrized
by a singular elliptic curve\begin{equation}
y^{2}=x^{3}-ax+b\qquad\mbox{with}\quad4a^{3}=27b^{2}\label{eq:elliptic-curve}\end{equation}
 with coordinates $(x,y)\in\mathbb{R}^{2}$ and parameters $a,b\in\mathbb{R}$.
Without loss of generality, we specialize to the concrete case $a=3,b=2$
with the double point at $y=0,x=1$. The tubular neighborhood of this
curve can be simply written as the complexification of the above curve,
i.e. $(x,y)\in\mathbb{C}^{2}$ where the double point is now located
along a disk centered at $y=0,x=1$. The double point defines a branch
point of index $2$.

\subsection*{{Appendix D - Spinor representation of immersed disks} }

{In this appendix we will follow the paper \cite{Friedrich1998}
very closely. Given a 2-disk $D^{2}$ and a 4-manifold $M$. The map
$i:D^{2}\to M$ is called an immersion if the differential $di:TD^{2}\to TM$
is injective. It is known from singularity theory \cite{GuiPol:74}
that every map of a 2-manifold into a 4-manifold can be deformed to
an immersion, the immersion may not be an embedding i.e. the immersed
disk may have self-intersections. For the following discussion we
consider the immersion $D^{2}\to U\subset\mathbb{R}^{4}$ of the disk
into one chart $U$ of $M$. }

{For aimplicity, start with a toy model of an immersion of a surface
into the 3-dimensional Euclidean space. Let $f:M^{2}\to\mathbb{R}^{3}$
be a smooth map of a Riemannian surface with injective differential
$df:TM^{2}\to T\mathbb{R}^{3}$, i.e. an immersion. In the } {\emph{Weierstrass
representation}}{ one expresses a }{\emph{conformal minimal}}{
immersion $f$ in terms of a holomorphic function $g\in\Lambda^{0}$
and a holomorphic 1-form $\mu\in\Lambda^{1,0}$ as the integral\[
f=Re\left(\int(1-g^{2},i(1+g^{2}),2g)\mu\right)\ .\]
 An immersion of $M^{2}$ is conformal if the induced metric $g$
on $M^{2}$ has components\[
g_{zz}=0=g_{\bar{z}\bar{z}}\,,\: g_{z\bar{z}}\not=0\]
 and it is minimal if the surface has minimal volume. Now we consider
a spinor bundle $S$ on $M^{2}$ (i.e. $TM^{2}=S\otimes S$ as complex
line bundles) and with the splitting\[
S=S^{+}\oplus S^{-}=\Lambda^{0}\oplus\Lambda^{1,0}\]
 Therefore the pair $(g,\mu)$ can be considered as spinor field $\varphi$
on $M^{2}$. Then the Cauchy-Riemann equation for $g$ and $\mu$
is equivalent to the Dirac equation $D\varphi=0$. The generalization
from a conformal minimal immersion to a conformal immersion was done
by many authors (see the references in\cite{Friedrich1998}) to show
that the spinor $\varphi$ now fulfills the Dirac equation\begin{equation}
D\varphi=K\varphi\label{eq:conformal-immersion-Dirac}\end{equation}
 where $K$ is the mean curvature (i.e. the trace of the second fundamental
form). The minimal case is equivalent to the vanishing mean curvature
$H=0$ recovering the equation above. Friedrich \cite{Friedrich1998}
uncovered the relation between a spinor $\Phi$ on $\mathbb{R}^{3}$
and the spinor $\varphi=\Phi|_{M^{2}}$: if the spinor $\Phi$ fulfills
the Dirac equation $D\Phi=0$ then the restriction $\varphi=\Phi|_{M^{2}}$
fulfills equation (\ref{eq:conformal-immersion-Dirac}) and $|\varphi|^{2}=const$.
Therefore we obtain\begin{equation}
H=\bar{\varphi}D\varphi\label{eq:mean-curvature-surface}\end{equation}
 with $|\varphi|^{2}=1$. }

{Now we will discuss the more complicated case. For that purpose
we consider the kinky handle which can be seen as the image of an
immersion $I:D^{2}\times D^{2}\to\mathbb{R}^{4}$. This map determines
a restriction of the immersion $I|_{\partial}:\partial D^{2}\times D^{2}\to\mathbb{R}^{4}$
with image a knotted solid torus $T(K)=I|_{\partial}(\partial D^{2}\times D^{2})$.
But a knotted solid torus $T(K)=K\times D^{2}$ is uniquely determined
by its boundary $\partial T(K)=K\times\partial D^{2}=K\times S^{1}$,
a knotted torus given as image $\partial T(K)=I_{\partial\times\partial}(T^{2})$
of the immersion $I|_{\partial\times\partial}:T^{2}=S^{1}\times S^{1}\to\mathbb{R}^{3}$.
But as discussed above, this immersion $I|_{\partial\times\partial}$
can be defined by a spinor $\varphi$ on $T^{2}$ fulfilling the Dirac
equation\begin{equation}
D\varphi=K\varphi\label{eq:2D-Dirac}\end{equation}
 with $|\varphi|^{2}=1$ (or an arbitrary constant) (see Theorem 1
of \cite{Friedrich1998}). The transition to the case of the immersion
$I|_{\partial}$ can be done by constructing a spinor $\phi$ out
of $\varphi$ which is constant along the normal of the immersed torus
$T^{2}$. As discussed above a spinor bundle over a surface splits
into two sub-bundles $S=S^{+}\oplus S^{-}$ with the corresponding
splitting of the spinor $\varphi$ in components\[
\varphi=\left(\begin{array}{c}
\varphi^{+}\\
\varphi^{-}\end{array}\right)\]
 and we have the Dirac equation\[
D\varphi=\left(\begin{array}{cc}
0 & \partial_{z}\\
\partial_{\bar{z}} & 0\end{array}\right)\left(\begin{array}{c}
\varphi^{+}\\
\varphi^{-}\end{array}\right)=K\left(\begin{array}{c}
\varphi^{+}\\
\varphi^{-}\end{array}\right)\]
 with respect to the coordinates $(z,\bar{z})$ on $T^{2}$. In dimension
3 we have a spinor bundle of same fiber dimension then the spin bundle
$S$ but without a splitting into two sub-bundles. Now we define the
extended spinor $\phi$ over the solid torus $\partial D^{2}\times D^{2}$
via the restriction $\phi|_{T^{2}}=\varphi$. Then $\phi$ is constant
along the normal vector $\partial_{N}\phi=0$ fulfilling the 3-dimensional
Dirac equation\begin{equation}
D^{3D}\phi=\left(\begin{array}{cc}
\partial_{N} & \partial_{z}\\
\partial_{\bar{z}} & -\partial_{N}\end{array}\right)\phi=K\phi\label{eq:Dirac-equation-3D}\end{equation}
 induced from the Dirac equation (\ref{eq:2D-Dirac}) via restriction
and where $|\phi|^{2}=const.$ Especially we obtain for the mean curvature\begin{equation}
K=\bar{\phi}D^{3D}\phi\label{eq:mean-curvature-3D}\end{equation}
 of the knotted solid torus $T(K)$ (up to a constant from $|\phi|^{2}$).
Or in local coordinates \begin{equation}
K=\overline{\phi}\sigma^{\mu}D_{\mu}^{3D}\phi\label{eq:trace-second-fund-form}\end{equation}
 with the Pauli matrices $\sigma^{\mu}$.}

\subsection*{{Appendix E - Extension of the action from 3D to 4D}}

{Now we will discuss the extension from the 3D to the 4D case. Let
$\iota:D^{2}\times S^{1}\hookrightarrow M$ be an immersion of the
solid torus $\Sigma=D^{2}\times S^{1}$ into the 4-manifold $M$ with
the normal vector $\vec{N}$. The spin bundle $S_{M}$ of the 4-manifold
splits into two sub-bundles $S_{M}^{\pm}$ where one subbundle, say
$S_{M}^{+},$ can be related to the spin bundle $S_{\Sigma}$. Then
the spin bundles are related by $S_{\Sigma}=\iota^{*}S_{M}^{+}$ with
the same relation $\phi=\iota_{*}\Phi$ for the spinors ($\phi\in\Gamma(S_{\Sigma})$
and $\Phi\in\Gamma(S_{M}^{+})$). Let $\nabla_{X}^{M},\nabla_{X}^{\Sigma}$
be the covariant derivatives in the spin bundles along a vector field
$X$ as section of the bundle $T\Sigma$. Then we have the formula\begin{equation}
\nabla_{X}^{M}(\Phi)=\nabla_{X}^{\Sigma}\phi-\frac{1}{2}(\nabla_{X}\vec{N})\cdot\vec{N}\cdot\phi\label{eq:covariant-derivative-immersion}\end{equation}
 with the obvious embedding $\phi\mapsto\left(\begin{array}{c}
\phi\\
0\end{array}\right)=\Phi$ of the spinor spaces. The expression $\nabla_{X}\vec{N}$ is the
second fundamental form of the immersion with trace the mean curvature
$2K$. Then from (\ref{eq:covariant-derivative-immersion}) one obtains
a similar relation between the corresponding Dirac operators\begin{equation}
D^{M}\Phi=D^{3D}\phi-K\phi\label{eq:relation-Dirac-3D-4D}\end{equation}
 with the Dirac operator $D^{3D}$ defined via (\ref{eq:Dirac-equation-3D}).
Together with equation (\ref{eq:Dirac-equation-3D}) we obtain\begin{equation}
D^{M}\Phi=0\label{eq:Dirac-equation-4D}\end{equation}
 i.e. $\Phi$ is a parallel spinor. }

{\emph{Conclusion:}}{ There is a relation between a 3-dimensional
spinor $\phi$ on a 3-manifold $\Sigma$ fulfilling a Dirac equation
$D^{\Sigma}\phi=K\phi$ (determined by the immersion $\Sigma\to M$
into a 4-manifold $M$) and a 4-dimensional spinor $\Phi$ on a 4-manifold
$M$ with fixed chirality ($\in\Gamma(S_{M}^{+})$ or $\in\Gamma(S_{M}^{-})$)
fulfilling the Dirac equation $D^{M}\Phi=0$. }

{From the Dirac equation (\ref{eq:Dirac-equation-4D}) we obtain
the the action\[
\intop_{M}\bar{\Phi}D^{M}\Phi\sqrt{g}\: d^{4}x\]
 as an extension of (\ref{eq:action-3D}) to the whole 4-manifold
$M$. By variation of the action (\ref{eq:action-3D}) we obtain an
immersion of minimal mean curvature, i.e. $K=0$. Then we can identify
via relation (\ref{eq:relation-Dirac-3D-4D}) the 4-dimensional and
the 3-dimensional action via \[
\intop_{M}\bar{\Phi}D^{M}\Phi\sqrt{g_{M}}\: d^{4}x=\intop_{T(K)}\bar{\phi}D^{3D}\phi\:\sqrt{g_{\partial}}\, d^{3}x=\intop_{T(K)}K\sqrt{g_{\partial}}\, d^{3}x\]
 Therefore the 3-dimensional action (\ref{eq:action-3D}) can be extended
to the whole 4-manifold (but for a spinor $\Phi$ of fixed chirality).
Finally we showed that the spinor can be extended to the whole 4-manifold
$M$.}


\begin{thebibliography}{10}
\bibitem{Akb:96} S.~Akbulut. \newblock Lectures on {S}eiberg-{W}itten
invariants. \newblock {\em Turkish J. Math.}, \textbf{20}:95--119,
1996.

\bibitem{AkbKir:79} S.~Akbulut and R.~Kirby. \newblock Mazur manifolds.
\newblock {\em Mich. Math. J.}, \textbf{26}:259--284, 1979.

\bibitem{Akbulut08} S.~Akbulut and K.~Yasui. \newblock Corks,
plugs and exotic structures. \newblock {\em Journal of Gokova Geometry
Topology}, 2:40--82, 2008. \newblock arXiv:0806.3010.

\bibitem{Akbulut09} S.~Akbulut and K.~Yasui. \newblock Knotted
corks. \newblock {\em J Topology}, 2:823--839, 2009. \newblock
arXiv:0812.5098.

\bibitem{Ashtekar08} A.~Ashtekar, J.~Engle, and D.~Sloan. \newblock
Asymptotics and {H}amiltonians in a first order formalism. \newblock
{\em Class. Quant. Grav.}, 25:095020, 2008. \newblock arXiv:0802.2527.

\bibitem{Ashtekar08a} A.~Ashtekar and D.~Sloan. \newblock Action
and {H}amiltonians in higher dimensional general relativity: First
order framework. \newblock {\em Class.Quant.Grav.}, 25:225025,
2008. \newblock arXiv:0808.2069.

\bibitem{AsselmeyerRose2010} T.~Asselmeyer-Maluga and H.~Ros{�}.
\newblock On the geometrization of matter by exotic smoothness. \newblock
arXiv:{[}gr-qc] 1006.2230, 2010.

\bibitem{Biz:94a} Z.~Bizaca. \newblock A handle decomposition of
an exotic {${\mathbb{R}}^{4}$}. \newblock {\em J. Diff. Geom.},
39:491 -- 508, 1994.

\bibitem{Biz:94} Z.~Bizaca. \newblock A reimbedding algorithm for
{C}asson handles. \newblock {\em Trans. Amer. Math. Soc.}, \textbf{345}:435--510,
1994.

\bibitem{Biz:95} Z.~Bizaca. \newblock An explicit family of exotic
{C}asson handles. \newblock {\em Proc. AMS}, 123:1297 -- 1302,
1995.

\bibitem{BizGom:96} {\u{Z}}.~Bi{\u{z}}aca and R~Gompf. \newblock
Elliptic surfaces and some simple exotic {${\Bbb{R}}^{4}$}'s. \newblock
{\em J. Diff. Geom.}, \textbf{43}:458--504, 1996.

\bibitem{Cas:73} A.~Casson. \newblock {\em Three lectures on
new infinite constructions in 4-dimensional manifolds}, volume~62.
\newblock Birkh{�}user, progress in mathematics edition, 1986.
\newblock Notes by Lucian Guillou, first published 1973.

\bibitem{CuFrHsSt:97} C.~Curtis, M.~Freedman, W.-C. Hsiang, and
R.~Stong. \newblock A decomposition theorem for h-cobordant smooth
simply connected compact 4-manifolds. \newblock {\em Inv. Math.},
\textbf{123}:343--348, 1997.

\bibitem{DonKro:90} S.~Donaldson and P.~Kronheimer. \newblock
{\em The Geometry of Four-Manifolds}. \newblock Oxford Univ. Press,
Oxford, 1990.

\bibitem{Fre:82} M.H. Freedman. \newblock The topology of four-dimensional
manifolds. \newblock {\em J. Diff. Geom.}, \textbf{17}:357 --
454, 1982.

\bibitem{Friedrich1998} T.~Friedrich. \newblock On the spinor representation
of surfaces in euclidean 3-space. \newblock {\em J. Geom. and Phys.},
28:143--157, 1998. \newblock arXiv:dg-ga/9712021v1.

\bibitem{GibHaw1977} G.W. Gibbons and S.W. Hawking. \newblock Action
integrals and partition functions in quantum gravity. \newblock {\em
Phys. Rev. D}, 15:2752--2756, 1977.

\bibitem{Gom:84} R.~Gompf. \newblock Infinite families of {C}asson
handles and topological disks. \newblock {\em Topology}, \textbf{23}:395--400,
1984.

\bibitem{Gom:89} R.~Gompf. \newblock Periodic ends and knot concordance.
\newblock {\em Top. Appl.}, \textbf{32}:141--148, 1989.

\bibitem{GomSti:1999} R.E. Gompf and A.I. Stipsicz. \newblock {\em
4-manifolds and {K}irby {C}alculus}. \newblock American Mathematical
Society, 1999.

\bibitem{GuiPol:74} V.~Guillemin and A.~Pollack. \newblock {\em
Differential topology}. \newblock Prentice-Hall, 1974.

\bibitem{Kato2004} T.~Kato. \newblock {ASD} moduli space over
four-manifolds with tree-like ends. \newblock {\em Geom. Top.},
8:779 -- 830, 2004. \newblock arXiv:math.GT/0405443.

\bibitem{Tau:87} C.H. Taubes. \newblock Gauge theory on asymptotically
periodic 4-manifolds. \newblock {\em J. Diff. Geom.}, \textbf{25}:363--430,
1987.

\bibitem{York1972} J.W. York. \newblock Role of conformal three-geometry
in the dynamics of gravitation. \newblock {\em Phys. Rev. Lett.},
28:1082--1085, 1972.
\end{thebibliography}
\end{document}